\documentclass[journal,comsoc]{IEEEtran}

\IEEEoverridecommandlockouts
\usepackage{cite}
\usepackage{amsmath,amssymb,amsfonts}
\usepackage{algorithmic}
\usepackage{graphicx}
\usepackage{textcomp}
\usepackage{xcolor}

\usepackage[T1]{fontenc}
\graphicspath{{myfigures/}}
\usepackage{epstopdf}					
\usepackage{amsfonts}
\usepackage{amssymb}
\usepackage{multirow}

\usepackage{color,soul}

\usepackage{bm}
\newtheorem{lem}{Lemma}
\newtheorem{theorem}{Theorem}[section]
\newcommand{\nid}{\noindent}
\newcommand{\argmin}{\operatornamewithlimits{argmin}}
\newcommand{\argmax}{\operatornamewithlimits{argmax}}

\usepackage{color}

\ifCLASSINFOpdf
\else
\fi
\usepackage{amsmath}
\usepackage[cmintegrals]{newtxmath}

\usepackage[belowskip=4pt,aboveskip=4pt,small]{caption}

\begin{document}
%
\title{Low-Complexity Detection for Faster-than-Nyquist Signaling based on Probabilistic Data Association}
%
%
%

\author{Michel~Kulhandjian,~\IEEEmembership{Member,~IEEE,
} Ebrahim~Bedeer,~\IEEEmembership{Member,~IEEE,
} Hovannes~Kulhandjian,~\IEEEmembership{Member,~IEEE,
} Claude~D'Amours,~\IEEEmembership{Member,~IEEE,
} and Halim~Yanikomeroglu,~\IEEEmembership{Fellow,~IEEE
}
\thanks{M. Kulhandjian and C. D'Amours are with the School of Electrical Engineering, \& Computer Science, University of Ottawa, Ottawa, Canada, e-mail: mkk6@buffalo.edu, cdamours@uottawa.ca.}
\thanks{E. Bedeer is with the Department of Electrical \& Computer Engineering, University of Saskatchewan, Saskatoon, Canada, e-mail: \mbox{e.bedeer@usask.ca}.}
\thanks{H. Kulhandjian is with the Department of Electrical \& Computer Engineering, California State University, Fresno, U.S.A., e-mail: \mbox{hkulhandjian@csufresno.edu}.}
\thanks{H. Yanikomeroglu is with the Department of Systems \& Computer Engineering, Carleton University, Ottawa, Canada, e-mail: halim@sce.carleton.ca.}

}

%
%

\markboth{IEEE Communications Letters, accepted}%
{Shell \MakeLowercase{\textit{et al.}}: Bare Demo of IEEEtran.cls for IEEE Communications Society Journals}
%



\maketitle

\begin{abstract}
In this paper, we investigate the sequence estimation problem of faster-than-Nyquist (FTN) signaling as a promising approach for increasing spectral efficiency (SE) in future communication systems. In doing so, we exploit the concept of Gaussian separability and propose two probabilistic data association (PDA) algorithms with polynomial time complexity to detect binary phase-shift keying (BPSK) FTN signaling. Simulation results show that the proposed PDA algorithm outperforms the
recently proposed SSSSE and SSSgb$K$SE algorithms for all SE values with a modest increase in complexity. The PDA algorithm approaches the performance of the semidefinite relaxation (SDRSE) algorithm for SE values of $0.96$ bits/sec/Hz, and it is within the $0.5$ dB signal-to-noise ratio (SNR) penalty at SE values of $1.10$ bits/sec/Hz for the fixed values of $\beta = 0.3$.
\end{abstract}


%
\IEEEpeerreviewmaketitle

\section{Introduction}
%
%
%
%

\IEEEPARstart{I}{mproving} spectral efficiency (SE) is one of the key requirements of future communication systems. Faster-than-Nyquist (FTN) signaling  is a promising physical layer transmission technique for increasing SE of future communication systems compared to conventional Nyquist signaling \cite{anderson2013faster}. The basic idea of FTN signaling is to transmit the time domain pulses with a rate that exceeds the Nyquist limit, which intentionally creates inter-symbol interference (ISI) at the receiver side. Interestingly, J. Mazo in 1975 showed that transmitting time domain pulses every $0.802\:T$, where $T$ is the symbol duration, results in the same minimum Euclidean distance (and, hence, the same error probability) as conventional Nyquist signaling \cite{mazo1975faster}. This is equivalent to a $25\%$ increase in SE for the same energy per symbol and transmission bandwidth. The concept of FTN signaling has been extended to different domains, e.g., frequency-domain \cite{rusek2005two}, non-binary modulation \cite{rusek2008non}, other pulse shapes \cite{liveris2003exploiting}, to name a few. 
Recent research results extended the concept of FTN signaling to different domains and suggested that it can be a promising candidates for high capacity peer-to-peer links.


Given that optimal detection of FTN signaling is shown to be NP-hard \cite{Lupas1989}, several linear/nonlinear detection techniques have been developed to remove ISI with different degrees of computational complexity \cite{anderson2013faster}. For high spectral efficiency systems, i.e., severe ISI, M-Bahl, Cocke, Jelinek, and Raviv (M-BCJR) algorithm \cite{anderson2009new} and the truncated Viterbi algorithm \cite{prlja2008receivers} are examples of nonlinear detectors that have shown SE gains; however, this has included exponential complexity in ISI length. {It is worth mentioning that iterative detection of FTN signaling was shown to be effective in approaching the zero-ISI performance at high signal-to-noise ratio (SNR) values in \cite{anderson2013faster, barbieri2009time}.}
To achieve moderate SE gains with reasonable detection complexity, nonlinear FTN signaling detection algorithms based on semi-definite relaxation were proposed in \cite{bedeer2017low} for high-order modulations with polynomial time complexity. A low-complexity nonlinear frequency domain equalizer was proposed in \cite{sugiura2013frequency}; however, it is found that the insertion of the cyclic prefix could result in performance loss for short block transmissions. 
The authors in \cite{bedeer2016very} proposed a linear algorithm that exploits the structure of ISI at the receiver and detects FTN signaling on a symbol-by-symbol basis. It was shown in \cite{bedeer2016very} that the proposed algorithm is suitable for low ISI scenarios.

{In this paper, we exploit the concept of Gaussian separability to propose {relatively} low-complexity probabilistic data association (PDA) algorithms to detect binary phase-shift keying (BPSK)}\footnote{The proposed algorithms can be straightforwardly extended to quadrature-phase-shift keying (QPSK).} {FTN signaling at polynomial time complexity.} The proposed algorithms are iterative, i.e., at each iteration a given FTN signaling symbol is detected to improve the overall error performance. Simulation results show the effectiveness of the proposed algorithms and their merit over other competing schemes in the literature. The proposed algorithms outperform the successive symbol-by-symbol sequence estimation (SSSSE) and successive symbol-by-symbol with go-back-$K$ sequence estimation (SSSgb$K$SE) algorithms in \cite{bedeer2016very} for all SE values. Additionally, the proposed algorithms approach the performance of the semidefinite relaxation (SDRSE) algorithm proposed in \cite{bedeer2017low} for SE values of $0.96$ bits/sec/Hz, and they are within $0.5$ dB SNR penalty at SE values of $1.10$ bits/sec/Hz. 


The remainder of this paper is organized as follows. Section~\ref{model} presents the FTN signaling system model and the linear separability concept. The concept of Gaussian separability and the proposed PDA algorithms are presented in Section \ref{proposed}, while the simulation results are discussed in Section \ref{simulation}. Finally, the paper is concluded in Section \ref{concluusion}.

The following notations are used in this paper. All boldface lower case letters indicate column vectors, and upper case letters indicate matrices. Additionally, $(\cdot)^H$ denotes Hermitian operation, $\mathsf{sgn}(\cdot)$ denotes the sign function, $| \cdot |$ is the scalar magnitude, $\mathsf{var}[\cdot]$ is variance, $\mathsf{cov}[\cdot]$ denotes covariance, $\mathsf{diag}[\cdot]$ denotes a matrix with diagonal elements, and $\mathbb{E} \{ \cdot \}$ denotes the expected value.

\section{FTN Signaling System Model}
\label{model}
At the FTN signaling transmitter, the BPSK data symbols are shaped by a unit-energy pulse $p(t)$. We transmit a total of $N$ symbols at every $\tau T$, where $\tau \in (0,1]$ is the time packing parameter.
The received signal, after being affected by an additive white Gaussian noise (AWGN) channel, passes through a filter matching the transmit pulse and is given as
\begin{eqnarray}
{y}(t) = \sqrt{\tau \: E_s} \: \sum\nolimits_{n = 1}^{N} {a}_n g(t - n \tau T) + {q}(t),
\end{eqnarray}
where ${a}_n, \: \: n= 1, \hdots, N,$ is the BPSK data symbol, $E_s$ is the data symbol  energy, $g(t) = \int\nolimits p(x) p(x - t) dx$, ${q}(t) = \int\nolimits n(x) p(x - t) dx$, where $n(t)$ is the AWGN with zero mean and variance $\sigma^2$, and $1/(\tau T)$ is the signaling rate. 
The output signal of the matched-filter is sampled at every $\tau T$ and can be expressed as 
\begin{eqnarray}\label{eq:ISI_initial}
{y}_k 
&=& \sqrt{\tau \:E_s} \sum\nolimits_{n = 1}^{N} {a}_n g(k \tau T - n \tau T) + {q}(k \tau T), \nonumber \\
&=& \underbrace{\sqrt{\tau \:E_s} \: {a}_k \: g(0)}_{\textup{desired symbol}}  \\ & & + \underbrace{\sqrt{\tau \: E_s} \: \sum\nolimits_{n = 1, \: n \ne k}^{N} {a}_n \: g((k - n) \tau T)}_{\textup{ISI}}  + \: {q}(k \tau T). \nonumber \\ \nonumber
\end{eqnarray}
\\[-0.6cm]
We rewrite  \eqref{eq:ISI_initial} in a vector form as 
\begin{eqnarray}\label{eq:vector_RX}
{\mathbf{y}}_{\textup{c}} &=& \sqrt{\tau \: E_s} \: {\mathbf{a}} \ast \mathbf{g} + {\mathbf{q}}_{\textup{c}}, 
\end{eqnarray} 
{{where $\mathbf{a}$, $\mathbf{g}$, $\mathbf{q}_c$, and $\ast$ are the transmit data symbol vector, ISI vector, colored noise vector, and convolution operator, respectively.}}
Since the noise samples are correlated, we design a whitening matched filter by
applying spectral factorization to the z-transform of $\mathbf{g}$ to obtain the approximate whitening filter coefficients. For more details on the design of such filters for FTN receivers, we refer the reader to  \cite{prlja2008receivers, bedeer2017low}.
Hence, after passing \eqref{eq:vector_RX} through the whitening filter, we have
\begin{eqnarray}\label{eq:complex}
{\mathbf{y}} &=& \sqrt{\tau \: E_s} \: {\mathbf{a}} \ast \mathbf{v} + {\mathbf{n}_{\rm{w}}}, 
\end{eqnarray}  
where ${\mathbf{n}}_{\rm{w}}$ is AWGN $\sim \mathcal{N}(0,\sigma^2)$ 
and $\mathbf{v}$ is the causal ISI vector and  is constructed from $\mathbf{g}$ as  $\mathbf{v}[n] \ast \mathbf{v}[-n] = \mathbf{g}$. Equation \eqref{eq:complex} can be rewritten  as 
\begin{equation}
\label{system01}
    {\mathbf{y}} = \mathbf{G}\mathbf{a} + {\mathbf{n}_{\rm{w}}},
\end{equation}
where $\mathbf{G}$ is an $N \times N$ Toeplitz Gram matrix that represents the ISI. 
The maximum likelihood sequence estimation (MLSE) problem for detecting the FTN signaling in \eqref{system01} is formally expressed as 
\begin{eqnarray}
\label{optimalML}
\widehat{\bm{a}} 
 &=&	\argmin_{\mathbf{a} \in \{\pm 1\}^{N \times 1}} \mathbf{a}^H \mathbf{G}\mathbf{a} - 2{\mathbf{y}}^H\mathbf{a}.
\end{eqnarray}
It is known that obtaining the MLSE solution is generally NP-hard \cite{Lupas1989}, which may be prohibitively complex for use in practical detectors. That being said, finding low-complexity optimal detectors of FTN signaling is important for improving SE in future communication systems.

In this work, we aim to identify what properties the ISI matrix $\mathbf{G}$ needs to have such that the sequence estimation can achieve  asymptotically zero probability of error detection.
In \cite{bedeer2016very}, the authors exploit the idea of linear separability to identify an operating region where perfect reconstruction of FTN signaling is guaranteed for noise-free transmission. It was found that such a linear separability operating region depends on the raised cosine pulse shape, its roll-off factor $\beta$, and the FTN signaling time acceleration parameter $\tau$. The perfect reconstruction conditions of the BPSK FTN signaling in \cite{bedeer2016very} can be summarized in the following lemma. 
 \begin{lem}
 \label{EbrahimLemma}
 \cite{bedeer2016very} We assume perfect estimation conditions for BPSK FTN signaling over the noise-free transmission. Regardless of the value of the current data symbol $a_k$, the upcoming $L-1$ data symbols $a_{k+1}, ..., a_{k+L-1}$, and the value of $L$, the following inequality holds for a certain range
 of $\tau$ and $\beta$:
 \vspace{-0.3cm}
 \begin{equation}
 \label{eq:lemma1}
     |G_{1,1} a_k | >  |G_{1,2} a_{k+1}+ \ldots + G_{1,L} a_{k+L-1}|,
 \end{equation}
 \end{lem}
 \noindent where $G_{i,j}$ is the $i$-th row and $j$-th column value of the ISI matrix $\mathbf{G}$.
Following \eqref{eq:lemma1}, in this work, we define the {\textit{conditionally linear margin}} as
\begin{equation}
    \label{deltak}
    \delta_k = |G_{1,1} a_k | -  |G_{1,2} a_{k+1}+ \ldots + G_{1,L} a_{k+L-1}|,
\end{equation}
which can be seen that $\delta_k>0$ is equivalent to \textit{Lemma} \ref{EbrahimLemma} and it is the sufficient and necessary condition for linear separability. Therefore, the linear margin, $\delta_k$, represents a distance measure for the separability condition of \textit{Lemma} \ref{EbrahimLemma} to hold.
Obviously, the greater the linear margin $\delta_k$, the better the error performance.

\section{Proposed Detection Algorithms}
\label{proposed}

In this section, low-complexity iterative detectors are proposed to detect FTN signaling on the basis of Gaussian separability \cite{Pattipati2001}. We show that Gaussian separability has a greater classification margin compared to the linear separability in \cite{bedeer2016very}, which thereby improves the detection error performance of FTN signaling.
To illustrate the idea, let us define a subset  $\mathcal{F} \subset \mathcal{U} = \{1, 2, \ldots, N\}$ of the symbols that are already detected, i.e., $\{a_l = a_l^*, \: l \in \mathcal{F}\}$. To detect the $k$-th symbol {$k \in \mathcal{U} - \mathcal{F}$}, we should be able to separate the set of undetected symbols in $\mathcal{U} - \mathcal{F}$ in two groups corresponding to $a_k = 1$ and $a_k = -1$, respectively. Following this notation, the observation vector ${\mathbf{y}}$ in (\ref{system01}) is rewritten as
\begin{eqnarray}
\label{system02}
{\mathbf{y}}     &=& \mathbf{g}_ka_k + \boldsymbol{\mu}_{\mathcal{F}}^* +  \boldsymbol{\mu}_{\mathcal{F}_k} + \mathbf{n}_{\rm{w}}, 
\end{eqnarray}
where $\mathbf{g}_k$ is the $k$-th column of ISI matrix $\mathbf{G}=[\mathbf{g}_1, \mathbf{g}_2, \dots, \mathbf{g}_N]$, $\boldsymbol{\mu}_{\mathcal{F}}^* = \sum_{l\in \mathcal{F}}\mathbf{g}_la_l^*$ and $ \boldsymbol{\mu}_{\mathcal{F}_k} = \sum_{j\in \mathcal{U} - \mathcal{F} -k}\mathbf{g}_ja_j$. 
 Linear detectors can be viewed as hyperplanes on the subspace spanned by $\mathbf{g}_k$ for $1 \leq k \leq N$ columns of matrix $\mathbf{G}$. A linear detector that achieves asymptotic negligible error probability as $\sigma^2 \rightarrow 0$ is said to be asymptotically efficient. Mathematically, it can be expressed as, $\forall a_j \in \{\pm 1 \}$, $j\in \mathcal{U} - \mathcal{F} -k$,
 \begin{align}
     \mathbf{c}^H \mathbf{y} &\geq 0 \: \: \text{if} \: \: a_k = 1 \nonumber \\
     \mathbf{c}^H \mathbf{y} &< 0 \: \: \text{if} \: \: a_k = -1. \nonumber
 \end{align}
 From this definition of linear separability, the following theorem can be proved.

 \begin{theorem}
 \label{TheoremLinear}
 \cite{Romano2005} The $k$-th symbol is asymptotically linearly separable conditionally on $\mathcal{F}$ if for $\sigma^2 \rightarrow 0$, an $N$-dimensional linear filter $\mathbf{c}$,
 \begin{equation}
 \label{condLinear}
 |\mathbf{c}^H \mathbf{g}_k| >  \sum_{j\in \mathcal{U} - \mathcal{F} -k}|\mathbf{c}^H \mathbf{g}_j |.
 \end{equation}
 \end{theorem}
{It can be shown that Lemma \ref{EbrahimLemma} is a special case of Theorem \ref{TheoremLinear} when the linear filter $\mathbf{c} = \mathbf{e}_k$ in (\ref{condLinear}), where $\mathbf{e}_k$ is a vector with value $1$ at position $k$ and $0$ in all other positions under the constraint that data symbols $a_k \in \{\pm 1\}$, $\forall k$.} 
\begin{table}[ht]
	\caption{Symbol-by-symbol Separability ($\beta = 0.3$) } 
	\centering 
		\begin{tabular}{c c c c c c} 
			\hline  
			\multicolumn{1}{c}{\multirow{2}{*}[-1.5pt]{$\tau$}} &
			\multicolumn{1}{c}{\multirow{2}{*}[-1.5pt]{0.6}}  & \multicolumn{1}{c}{\multirow{2}{*}[-1.5pt]{0.7}} & \multicolumn{1}{c}{\multirow{2}{*}[-1.5pt]{0.8}} & \multicolumn{1}{c}{\multirow{2}{*}[-1.5pt]{0.9}} \\  [2.6ex] 
			\hline   \rule{-3pt}{2.5ex} 
			 $\delta_{max}, \delta_{ave}$ & $0.97, 0.11$  & $0.97, 0.20$ & $0.97, 0.38$  & $0.97, 0.53$\\[0.6ex]
			\hline 
		\end{tabular} \label{linearS}
		\vspace{-0.2cm}
\end{table}
\begin{table}[ht]
	\caption{Gaussian Separability ($\beta = 0.3$) } 
	\centering 
		\begin{tabular}{c c c c c c} 
			\hline\hline  
			\multicolumn{1}{c}{\multirow{2}{*}[-1.5pt]{SNR (dB)}} &
			\multicolumn{1}{c}{\multirow{2}{*}[-1.5pt]{$\tau$ = 0.6}}  & \multicolumn{1}{c}{\multirow{2}{*}[-1.5pt]{0.7}} & \multicolumn{1}{c}{\multirow{2}{*}[-1.5pt]{0.8}} & \multicolumn{1}{c}{\multirow{2}{*}[-1.5pt]{0.9}} \\  [2.6ex] 
			\hline   \rule{-3pt}{2.5ex} 
			$0$ & $0.00, 0.00$  & $1.16, 0.49$ & $1.08, 0.58$  & $1.02, 0.73$\\[0.6ex]
			2 & $2.01, 0.68$    & $1.84, 0.82$ & $1.71, 0.96$ & $1.62, 1.21$ \\[0.6ex]
			4 & $3.19, 1.12$   & $2.92, 1.35$ & $2.71, 1.57$ & $2.56, 2.00$ \\[0.6ex]
			6 & $5.06, 1.73$   &$4.62, 2.16$ & $4.29, 2.55$ & $4.06, 3.30$ \\[0.6ex]
			8 & $8.02, 2.71$   &$7.32, 3.45$ & $6.80, 4.13$ & $6.43, 5.42$ \\[0.6ex]
			\hline 
		\end{tabular} \label{gaussianS}
		\vspace{-0.2cm}
\end{table}


Recall that if the condition densities for all the symbols $\mathbf{f}(\mathbf{y} | a_j )$ are Gaussian then the optimal filter for user-$k$ is a scaled version of e.g., $\mathbf{c}_k \triangleq \mathbf{R}_{\mathcal{F}_k}^{-1} \mathbf{g}_k$.
 The solution of the optimal separating hyperplane in a linear detector (classifier), despite the simple formulation, may be very difficult when the classes do not follow a Gaussian distribution. If we assume that the ISI is Gaussian, then the linear classifier becomes the optimal maximum likelihood (ML) decision with hyperplanes dividing the $N$-dimensional space.
\begin{theorem}
\cite{Romano2005} The $k$-th symbol is a Gaussian separable conditionally on $\mathcal{F}$ if for $\sigma^2 \rightarrow 0 $,
\begin{equation}
\label{condGaussian}
    \mathbf{g}_k^H \mathbf{R}_{\mathcal{F}_k}^{-1} \mathbf{g}_k >  \sum_{j\in \mathcal{U} - \mathcal{F} -k}|\mathbf{g}_k^H \mathbf{R}_{\mathcal{F}_k}^{-1} \mathbf{g}_j |,
\end{equation}
\\[-0.6cm]
\end{theorem}
where
 \begin{equation}
     \mathbf{R}_{\mathcal{F}_k} =  \sum_{j\in \mathcal{U} - \mathcal{F} -k} \mathbf{g}_j \mathbf{g}_j^H + \sigma^2 \mathbf{I}_N,
 \end{equation}
 \\[-0.3cm]
and the parameter
\begin{equation}
\label{DeltaGaussiank}
    \Delta_{\mathcal{F}_k} = \mathbf{g}_k^H \mathbf{R}_{\mathcal{F}_k}^{-1} \mathbf{g}_k -  \sum_{j\in \mathcal{U} - \mathcal{F} -k}|\mathbf{g}_k^H \mathbf{R}_{\mathcal{F}_k}^{-1} \mathbf{g}_j |
\end{equation}
is called {\textit{conditionally Gaussian margin}}. 
Unlike the linear margin $\delta_k$, the Gaussian margin $\Delta_{\mathcal{F}_k}$ value depends on the noise variance $\sigma^2$. {We measured conditionally margins, $\delta_k$ and $\Delta_{\mathcal{F}_k}$, in the operating region, where ISI matrix $\mathbf{G}$ is formed with roll-off factor $\beta = 0.3$ and $\tau \in \{0.6, 0.7, 0.8, 0.9 \}$. At each pair of $\beta, \tau$, we generate matrix $\mathbf{G}$ and then compute $\delta_k$ and {$\Delta_{\mathcal{F}_k}$} for $1 \leq k \leq N$ using equations (\ref{deltak}) and (\ref{DeltaGaussiank})\footnote{We note that $\Delta_{\mathcal{F}_k}$ is computed using the following set $\mathcal{F} = \{i_1,i_2, \dots, i_{k-1} \}$ for $1\leq k \leq N$. The order of $\{i_1,i_2, \dots, i_{k-1} \}$ depends on the ISI matrix $\mathbf{G}$ and $i_t$'s are precisely computed in our PDA algorithm at line $6$.}, respectively. We evaluate their maximum and average values as $\delta_{max} = \argmax_{k} \delta_k$, $\delta_{ave}=\frac{1}{N}\sum_{k = 1}^N\delta_k$, {$\Delta_{max} = \argmax_{k} \Delta_{\mathcal{F}_k}$}, $\Delta_{ave} = \frac{1}{N}\sum_{k = 1}^N\Delta_{\mathcal{F}_k}$.

Since the linear margin $\delta_k$ does not depend on the noise variance, $\sigma^2$, there is only one row with values ($\delta_{max}$, $\delta_{ave}$) as shown in Table \ref{linearS}. On the other hand, Table \ref{gaussianS} shows the ($\Delta_{max}$, $\Delta_{ave}$) for different SNR values at each row, where we vary the noise variance, $\sigma^2$, and keep the matrix $\mathbf{G}$ unchanged. We observe from Table \ref{linearS} and Table \ref{gaussianS} that the Gaussian classification margins have greater maximum and average values than their linear counterparts. Therefore, we can theorize that using a Gaussian classification margin can potentially improve the performance in terms of bit error rate (BER).}

The main idea of the PDA detection algorithm is based on conditionally Gaussian separability criterion and the fact that the Gaussian margin $\Delta_{\mathcal{F}_k}$ is greater than the linear margin $\delta_k$, as shown in Tables \ref{linearS} and \ref{gaussianS}. One can re-express \eqref{system02} as
\begin{eqnarray}
\label{system03}
  {\mathbf{y}}   &=& {\mathbf{g}_ka_k +  \mathbf{w}_k},  
\end{eqnarray}
where $\mathbf{w}_k = \mathbf{G}_k \mathbf{a}_k + \mathbf{n}_{\rm{w}}$, $\mathbf{G}_k$ is the $N\times N-1$ resulting matrix after excluding the $k$-th column of $\mathbf{G}$, and $\mathbf{a}_k$ is $N-1\times 1$ vector resulting after excluding $k$-th element of $\mathbf{a}$.

We associate the probabilities $P_a(k)$ and $1-P_a(k)$ with the estimate of the symbols $a_k = 1$ and $a_k = -1$, respectively.
Now, for an arbitrary symbol $a_k$, we treat the other symbols $a_j$, $\forall j \neq k$ as binary random variables and $\mathbf{w}_k$ as a Gaussian noise vector. Then the likelihood ratio for the $a_k$ symbol can be written as
\vspace{-0.0cm}
\begin{eqnarray}
\label{likelihood01}
  \Lambda(a_k)  &=&\frac{P\{a_k = 1 | \mathbf{y}, \{P_a(j) \}_{j \neq k} \}}{P\{a_k = -1 | \mathbf{y}, \{P_a(j) \}_{j \neq k} \} }, \nonumber \\ 
  &=&\frac{f_{\mathbf{y}}(\mathbf{y} | a_k = 1 | , \{P_a(j) \}_{j \neq k} )}{f_{\mathbf{y}}(\mathbf{y} | a_k = -1 , \{P_a(j) \}_{j \neq k} ) }\\ \nonumber
  &=&\frac{\mathcal{N}(\mathbf{y} | \mathbf{g}_k + \boldsymbol{\mu}_k, \mathbf{C}_k )}{\mathcal{N}(\mathbf{y} | \mathbf{g}_k - \boldsymbol{\mu}_k, \mathbf{C}_k ) }\\ 
    &=& \exp\left[2(\bm{y} - \boldsymbol{\mu}_k)^H\mathbf{C}_k^{-1}\mathbf{g}_k\right],
\end{eqnarray}
where $P_a(k) = P\{a_k = 1 | \mathbf{y}, \{P_a(j) \}_{j \neq k} \}$ and $f_{\mathbf{y}}$ is the multivariate Gaussian distribution conditioned on  $\{\mathbf{y}, \{P_a(j) \}_{j \neq k} \}$. The covariance can be calculated as
\begin{eqnarray}
\label{covariance01}
  \mathbf{C}_k  &=&\mathsf{cov}[\mathbf{w}_k]\nonumber \\ \nonumber
    &=& \mathbf{G}_k \mathbb{E}[\mathbf{a}_k\mathbf{a}_k^H]\mathbf{G}_k^H -  \mathbf{G}_k \mathbb{E}[\mathbf{a}_k]\mathbb{E}[\mathbf{a}_k^H]\mathbf{G}_k^H + \mathbf{R}_n \nonumber\\
    &=& \mathbf{G}_k \mathbf{C}_{\mathbf{a}_k}\mathbf{G}_k^H + \mathbf{R}_n,
\end{eqnarray} 
where $\mathsf{var}[a_k] = 4P_a(k)(1-P_a(k))$, $\mathbf{C}_{\mathbf{a}_k} = \mathsf{cov}[\mathbf{a}_k] = \mathsf{diag}(\mathsf{var}[a_1],\cdots, \mathsf{var}[a_{k-1}], \mathsf{var}[a_{k+1}], \cdots, \mathsf{var}[a_n])$ and $\mathbf{R}_n = \mathbb{E}[\mathbf{n}_{\rm{w}}\mathbf{n}_{\rm{w}}^H]$, are the covariance matrix of $\mathbf{a}_k$ and the autocorrelation of noise vector, respectively. The knowledge about the $a_k$ symbol is carried over in each iteration through a dynamic update of means and covariances as follows
 \begin{eqnarray}
 \label{mean01}
   \boldsymbol{\mu}_k  &=&\mathbb{E}[\mathbf{w}_k] = \mathbb{E}[\mathbf{G}_k\mathbf{a}_k + \mathbf{n}_{\rm{w}}]= \mathbf{G}_k\mathbb{E}[\mathbf{a}_k] \nonumber\\ \nonumber
     &=& \mathbf{G}_k\begin{bmatrix}
            2P_a(1) -1 \\
            \vdots \\
            2P_a(k-1) -1\\
            2P_a(k+1) -1\\
            \vdots \\
            2P_a(N) -1
          \end{bmatrix} \\
     &=& \mathbf{G}_k \mathbf{p}_{a,k},
\end{eqnarray}
where $\mathbf{p}_{a,k}$ is the estimated $\mathbf{a}$ signal vector without the $k$-th signal. As we can see, the likelihood ratio  in \eqref{likelihood01} is in the form of $P_a(k)/(1-P_a(k))= \exp(x)$. Hence, the posteriori probability $P_a(k)$ can be calculated as
\begin{eqnarray}\label{eq:ComCal}
P_a(k) = 1/\left(1 + \exp\left[-2(\mathbf{y}-\boldsymbol{\mu}_k)^H\mathbf{C}_k^{-1}\mathbf{g}_k\right]\right).
\end{eqnarray}


That being said, the proposed PDA algorithm for detecting FTN signaling can be summarized in the next page, where $M$ is the number of iterations and $D_k = \mathbf{g}_k^H \mathbf{C}_{k}^{-1}\mathbf{g}_k$.
 
\begin{table}
		\normalsize
		\centering  %
		\begin{tabular}{l}
			\hline \hline \rule{0pt}{3ex} 
			\nid \textbf{Proposed PDA Algorithm }  \\
			\hline \rule{0pt}{3ex} 
			\nid \textbf{{Input}:} $\mathbf{y}$, $\mathcal{F} \gets \{1,2, \dots, N\}$, $M$ \\
			\hspace{0.1cm} 1:\hspace{0.0cm} Initialize : $P_a(k) = 1/2$, $k = 1, \dots, N$ \\
			\hspace{0.1cm} 2:\hspace{0.0cm}  Compute $D_k$, $k = 1, \cdots, N$ \\ 
			\hspace{0.1cm} 3:\hspace{0.0cm} \textbf{for} $i = 1$ to $M$  \\
			\hspace{0.1cm} 4:\hspace{0.3cm}  $\mathcal{U} \gets \mathcal{F}$\\ 
			\hspace{0.1cm} 5:\hspace{0.3cm} \textbf{while} ($\mathcal{U} \neq \emptyset$) \\
			\hspace{0.1cm} 6:\hspace{0.6cm} Update $P_a(k_m)$, $k_m = \argmax_{k\in \mathcal{U}}(D_k)$\\
			\hspace{0.1cm} 7:\hspace{0.6cm} Subtract $k_m$ from $\mathcal{U}$   \\
			\hspace{0.1cm} 8:\hspace{0.6cm} Recompute $D_k$, $\forall k \in \mathcal{U}$ \\
			\nid \textbf{{Output}:} $\mathbf{\hat{a}} = \mathsf{sgn}\left ( \mathbf{p}_a \right )$ \\
			\hline
		\end{tabular}\vspace{-0.0cm}
		\label{originalPDA}
		\vspace{-0.4cm}
	\end{table}
	
\vspace{-0.5cm}
\subsection{Modified PDA algorithm}
The complexity of the PDA algorithm can be further reduced if the detected data symbols whose probabilities lie within a confidence interval are removed in each iteration. 
In other words, if the probability of the symbol $k$, $P_a(k)$ lies in confidence region $\epsilon$, if $|P_a(k)-d| <\epsilon$, where $d = \{0,1\}$. All the symbols that satisfy the confidence interval define the set $\mathcal{C}$. We assign $P_a(k)=\{d : |P_a(k)-d|<\epsilon\}$ for $\forall k \in \mathcal{C}$. Next, we remove symbols $\mathcal{C}$ from $\mathcal{F}$, $\mathcal{F} = \mathcal{F} - \mathcal{C}$. The modified PDA algorithm is presented next.

\begin{table}[ht]
	\centering  %
	\normalsize
	\begin{tabular}{l}
		\hline \hline \rule{0pt}{3ex} 
		\nid \textbf{Modified PDA Algorithm }  \\
		\hline \rule{0pt}{3ex} 
		\nid \textbf{{Input}:} $\mathbf{y} = 
		\mathbf{G}\mathbf{a} + \mathbf{n}$, $\mathcal{F} \gets \{1,2, \dots, N\}$, $M$ \\
		\hspace{0.1cm} 1:  Initialize $P_a(k) = 1/2$, $k = 1, \cdots, N$  \\
		\hspace{0.1cm} 2:  Compute $D_k$, $k = 1, \dots, N$\\
		\hspace{0.1cm} 3:  \textbf{for} $i = 1$ to $M$\\
		\hspace{0.1cm} 4: \hspace{0.3cm}  Find $k$ that satisfy  $|P_a(k)-d| <\epsilon$ \\
		\hspace{0.1cm} 5: \hspace{0.3cm} \textbf{if} $\mathcal{C} \neq \emptyset$\\
		\hspace{0.1cm} 6: \hspace{0.6cm}  $P_a(k)=d$, $\forall k \in \mathcal{C}$\\
		\hspace{0.1cm} 7: \hspace{0.6cm}  $\mathcal{F} \gets \mathcal{F} - \mathcal{C}$\\
		\hspace{0.1cm} 8:  \hspace{0.3cm} $\mathcal{U} \gets \mathcal{F}$ \\
		\hspace{0.1cm} 9:  \hspace{0.3cm} \textbf{while} ($\mathcal{U} \neq \emptyset$) \\
		\hspace{0.1cm}10:  \hspace{0.6cm} Update $P_a(k_m)$, $k_m = \argmax_{k\in \mathcal{U}}(D_k)$ \\
		\hspace{0.1cm}11:  \hspace{0.6cm} Subtract $k_m$ from $\mathcal{U}$  \\
		\hspace{0.1cm}12:  \hspace{0.6cm} Recompute $D_k$, $\forall k \in \mathcal{U}$  \\
		\nid \textbf{{Output}:} $\mathbf{\hat{a}}=\mathsf{sgn}\left ( \mathbf{p}_a \right )$ \\
		\hline
	\end{tabular}\vspace{-0.0cm}
	\label{modifiedPDA}
	\vspace{-0.5cm}
\end{table}
 
\subsection{Complexity analysis}
{The main complexity of the proposed PDA algorithm lies in updating $\mathbf{C}_k^{-1}$ in (19). {However, efficient numerical techniques such as the Durbin-Levinson algorithm can exploit the Toeplitz structure of $C_k$ and  compute its inverse, i.e., $\mathbf{C}_k^{-1}$, with a complexity of $\mathcal{O}(N^2)$ \cite{wysocki2013digital}}. However, since $C_k$ is not always guaranteed to be Toeplitz, the calculations of $C_k^{-1}$ can be in the order of $\mathcal{O}(N^3)$}. Since evaluating \eqref{eq:ComCal} is expected to occur at most $N$ times and the algorithm has deterministic execution steps, the total average and worst case computational complexity of the proposed PDA algorithm can be in the order of $\mathcal{O}(N^4)$. 
{The PDA algorithm has the same polynomial complexity order (i.e., both complexities scale in polynomial time with the number of transmit symbols) as the SDRSE algorithm in \cite{bedeer2017low} (please note that complexity measure based on the number of operations depends mainly on hardware implementations and is outside the scope of this work), while it has higher complexity compared to the work in \cite{bedeer2016very}.}
\vspace{-0.2cm}

\section{Simulation Results}
\label{simulation}
In this section, we assess the performance of our proposed algorithms for detecting BPSK FTN signaling with a root-raised cosine (rRC) pulse and roll-off factor of $\beta = 0.3$. Simulations were performed over the AWGN channel, both with and without channel encoding. {The iteration number for proposed PDA is $8$ for all cases and the number of iterations we averaged the simulations over are $50,000$ and $5,000$ for uncoded and coded cases, respectively.}

\begin{figure}[t]
\vspace{-.6cm}
	\centering
	\includegraphics[width=0.50\textwidth]{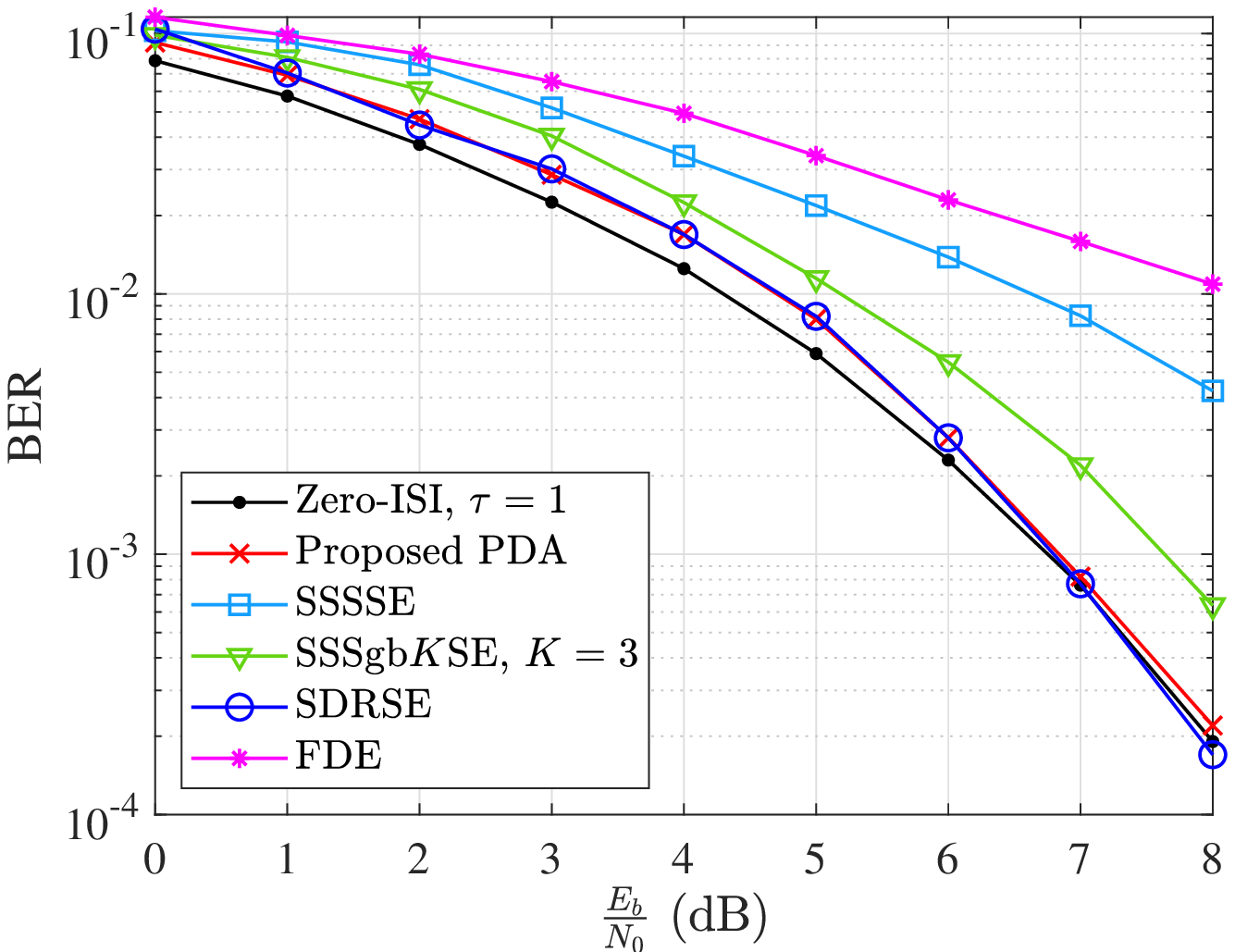}
	\caption{BER performance of {BPSK} FTN signaling as a function of $\frac{E_b}{N_o}$ using the proposed PDA algorithm, SSSSE \cite{bedeer2016very}, SSSgb$K$SE \cite{bedeer2016very}, and the SDRSE \cite{bedeer2017low}, at $\beta = 0.3$ and $\tau = 0.8$, SE $=0.96$ bits/sec/Hz.}\label{fig:tau_08}
	\vspace{-.4cm}
\end{figure}

{
Figure \ref{fig:tau_08} depicts the BER of BPSK FTN signaling as a function of $\frac{E_b}{N_o}$ for the proposed PDA algorithm, comparing it with the SSSE and SSSgb$K$SE ($K = 3$) algorithms presented in \cite{bedeer2016very}, the SDRSE algorithm in \cite{bedeer2017low}, and the frequency domain equalizer (FDE) in \cite{sugiura2013frequency} for $\beta = 0.3$ and $\tau = 0.8$, i.e., a SE of $0.96$ bits/sec/Hz. The SE is calculated as $\frac{\log_2{M}}{(1 + \beta)\tau}$, where $M$ is the constellation size and the SE loss due to adding the cyclic prefix in the FDE in \cite{sugiura2013frequency} is not considered. As we can see, the proposed PDA algorithm approaches the performance of the SDRSE algorithm in \cite{bedeer2017low} and outperforms the works in \cite{bedeer2016very, sugiura2013frequency} for SE value of $0.96$ bits/sec/Hz.}

\begin{figure}[!t]
\vspace{-.6cm}
	\centering
	\includegraphics[width=0.50\textwidth]{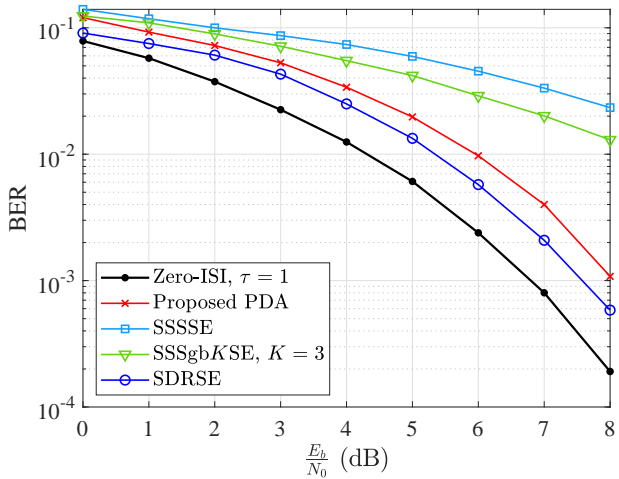}
	\caption{BER performance of {BPSK} FTN signaling as a function of $\frac{E_b}{N_o}$ using the proposed PDA algorithm, SSSSE \cite{bedeer2016very}, SSSgb$K$SE \cite{bedeer2016very}, and the SDRSE \cite{bedeer2017low}, at $\beta = 0.3$ and $\tau = 0.7$, SE $=1.10$ bits/sec/Hz.}\label{fig:tau_07}
		\vspace{-.2cm}
\end{figure}
\vspace{-0.0cm}
 \begin{figure}[!t]
 	\vspace{-.3cm}
	\centering
	\includegraphics[width=0.50\textwidth]{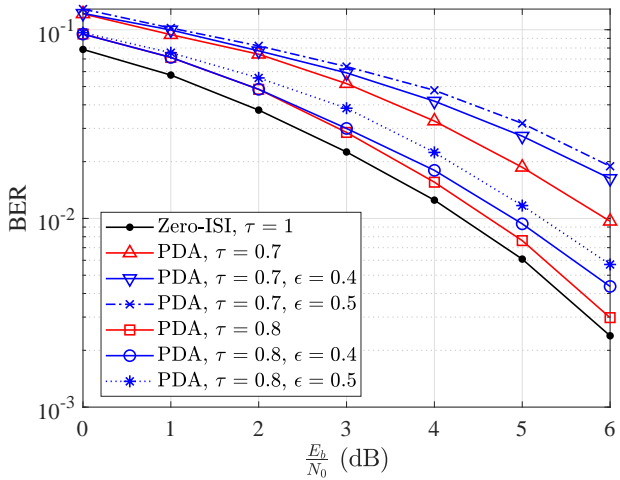}
	\caption{BER performance for confidence region.}\label{fig:confidenceR_01}
	\vspace{-.6cm}
\end{figure}
Figure \ref{fig:tau_07} plots the BER of the BPSK FTN signaling as a function of $\frac{E_b}{N_o}$ for the proposed PDA algorithm, the SSSSE and SSSgb$K$SE ($K = 3$) algorithms in \cite{bedeer2016very}, and the SDRSE algorithm in \cite{bedeer2017low} for $\beta = 0.3$ and $\tau = 0.7$, i.e., a SE of $1.10$ bits/sec/Hz. As we can see in Fig. \ref{fig:tau_07}, the proposed PDA algorithm's performance is within almost $0.5$ dB of the SDRSE performance in \cite{bedeer2017low}, and it extensively outperforms the works in \cite{bedeer2016very} that fail at such value of SE. We observe in Fig. 3 of our simulation for the confidence region $\epsilon = 0.4$, the average number of iteration steps reduces by around $30\%$. At higher SNR values, we observe higher percentage of reduction in the iteration steps.

\begin{figure}[!t]
	\centering
	\includegraphics[width=0.50\textwidth]{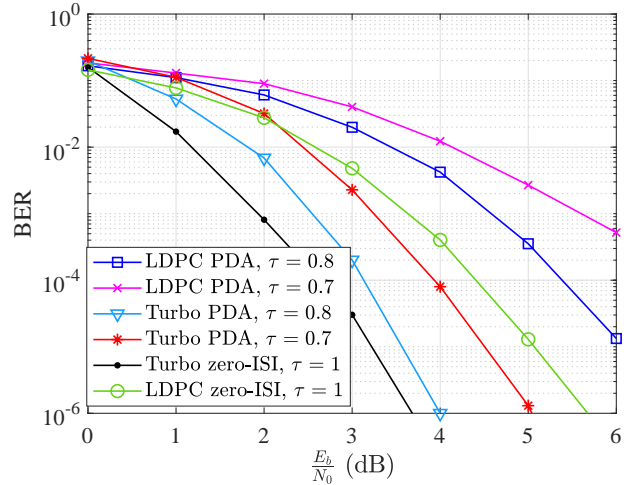}
	\caption{BER performance for LDPC and turbo encoding.}\label{fig:FTN_WMF17_coded_02}
		\vspace{-.6cm}
\end{figure}

 We also performed the simulation using a low-density parity-check (LDPC) and turbo codes. For implementation, we used a custom parity based matrix generator for LDPC and the long-term evolution (LTE) turbo channel codes described in \cite{Zarrinkoub2014}. For both cases, we used $1/3$ code rate channel coding with an input message length of $320$ and an output code length of $972$ for both LDPC and LTE turbo, respectively. 
 The construction of the LTE interleaver was based on the quadratic permutation polynomial (QPP) scheme \cite{Zarrinkoub2014}. 
 Figure \ref{fig:FTN_WMF17_coded_02} shows the BER performance for LDPC and turbo encoded with $\tau = 0.7$ and $\tau = 0.8$. The BER performance of turbo coded is better compared to LDPC in both cases. The SSSE and SSSgb$K$SE are not included in the simulation results since they do not produce soft outputs required by the LDPC and turbo soft decoders.
\vspace{-0.3cm}
 
\section{Conclusion}
\vspace{-0.0cm}
\label{concluusion}
Faster-than-Nyquist (FTN) signaling is a promising candidate for improving spectral efficiency (SE) in future communication systems. In this paper, we showed that Gaussian separability has a greater margin compared to linear separability, which has been used recently in the detection of FTN. In exploiting Gaussian separability, we proposed two probabilistic data association (PDA) algorithms of polynomial time computational complexity to detect binary phase-shift keying (BPSK) FTN signaling. Simulation results showed that the proposed algorithms outperform the successive symbol-by-symbol sequence estimation (SSSSE) and successive symbol-by-symbol with go-back-$K$ sequence estimation (SSSgb$K$SE) algorithms in \cite{bedeer2016very} for all SE values. Additionally, they approach the performance the semidefinite relaxation (SDRSE) algorithm in \cite{bedeer2017low} for SE values of $0.96$ bits/sec/Hz and they are within the $0.5$ dB signal-to-noise ratio (SNR) penalty at SE values of $1.10$ bits/sec/Hz. 

\bibliographystyle{IEEEtran}
\bibliography{IEEEabrv,mybib_file}

\begin{thebibliography}{10}
\providecommand{\url}[1]{#1}
\csname url@samestyle\endcsname
\providecommand{\newblock}{\relax}
\providecommand{\bibinfo}[2]{#2}
\providecommand{\BIBentrySTDinterwordspacing}{\spaceskip=0pt\relax}
\providecommand{\BIBentryALTinterwordstretchfactor}{4}
\providecommand{\BIBentryALTinterwordspacing}{\spaceskip=\fontdimen2\font plus
\BIBentryALTinterwordstretchfactor\fontdimen3\font minus
  \fontdimen4\font\relax}
\providecommand{\BIBforeignlanguage}[2]{{%
\expandafter\ifx\csname l@#1\endcsname\relax
\typeout{** WARNING: IEEEtran.bst: No hyphenation pattern has been}%
\typeout{** loaded for the language `#1'. Using the pattern for}%
\typeout{** the default language instead.}%
\else
\language=\csname l@#1\endcsname
\fi
#2}}
\providecommand{\BIBdecl}{\relax}
\BIBdecl

\bibitem{anderson2013faster}
J.~B. Anderson, F.~Rusek, and V.~{\"O}wall, ``{Faster-than-Nyquist
  signaling},'' \emph{Proc. {IEEE}}, vol. 101, no.~8, pp. 1817--1830, Aug.
  2013.

\bibitem{mazo1975faster}
J.~Mazo, ``{Faster-than-Nyquist signaling},'' \emph{Bell Syst. Tech. J.},
  vol.~54, no.~8, pp. 1451--1462, Oct. 1975.

\bibitem{rusek2005two}
F.~Rusek and J.~B. Anderson, ``{The two dimensional Mazo limit},'' in
  \emph{Proc. IEEE Int. Symp. Inf. Theory (ISIT)}, Adelaide, Australia, Sep.
  2005, pp. 970--974.

\bibitem{rusek2008non}
------, ``{Non binary and precoded faster than Nyquist signaling},''
  \emph{{IEEE} Trans. Commun.}, vol.~56, no.~5, pp. 808--817, May 2008.

\bibitem{liveris2003exploiting}
A.~D. Liveris and C.~N. Georghiades, ``{Exploiting faster-than-Nyquist
  signaling},'' \emph{{IEEE} Trans. Commun.}, vol.~51, no.~9, pp. 1502--1511,
  Sep. 2003.

\bibitem{Lupas1989}
R.~Lupas and S.~Verdu, ``{Linear multiuser detectors for synchronous
  code-division multiple-access channels},'' \emph{IEEE Trans. Inf. Theory},
  vol.~35, no.~1, pp. 123--136, Jan. 1989.

\bibitem{anderson2009new}
J.~B. Anderson, A.~Prlja, and F.~Rusek, ``New reduced state space {BCJR}
  algorithms for the {ISI} channel,'' in \emph{Proc. IEEE Int. Symp. Inf.
  Theory (ISIT)}, Seoul, Korea, Jun. 2009, pp. 889--893.

\bibitem{prlja2008receivers}
A.~Prlja, J.~B. Anderson, and F.~Rusek, ``{Receivers for faster-than-Nyquist
  signaling with and without turbo equalization},'' in \emph{Proc. IEEE Int.
  Symp. Inf. Theory (ISIT)}, Toronto, Canada, Jul. 2008, pp. 464--468.

\bibitem{barbieri2009time}
A.~{Barbieri}, D.~{Fertonani}, and G.~{Colavolpe}, ``{Time-frequency packing
  for linear modulations: spectral efficiency and practical detection
  schemes},'' \emph{IEEE Trans. Commun.}, vol.~57, no.~10, pp. 2951--2959, Oct.
  2009.

\bibitem{bedeer2017low}
E.~Bedeer, M.~H. Ahmed, and H.~Yanikomeroglu, ``Low-complexity detection of
  high-order {QAM} faster-than-{Nyquist} signaling,'' \emph{IEEE Access},
  vol.~5, pp. 14\,579--14\,588, Jul. 2017.

\bibitem{sugiura2013frequency}
S.~Sugiura, ``{Frequency-domain equalization of faster-than-Nyquist
  signaling},'' \emph{IEEE Wireless Commun. Lett.}, vol.~2, no.~5, pp.
  555--558, Oct. 2013.

\bibitem{bedeer2016very}
E.~Bedeer, M.~H. Ahmed, and H.~Yanikomeroglu, ``A very low complexity
  successive symbol-by-symbol sequence estimator for faster-than-{Nyquist}
  signaling,'' \emph{IEEE Access}, vol.~5, no.~1, pp. 2169--3536, Mar. 2017.

\bibitem{Pattipati2001}
J.~{Luo}, K.~R. {Pattipati}, and P.~K. {Willett}, ``{Near-optimal multiuser
  detection in synchronous CDMA using probabilistic data association},''
  \emph{IEEE Commun. Lett.}, vol.~5, no.~9, pp. 361--366, Sept. 2001.

\bibitem{Romano2005}
G.~{Romano}, F.~{Palmieri}, P.~K. {Willett}, and D.~Mattera, ``{Separability
  and gain control for overloaded CDMA},'' in \emph{Proc. of the Conf. on Inf.
  Sci. and Sys. (CISS)}, Princeton, NJ, USA, Mar. 2005, pp. 1--6.

\bibitem{wysocki2013digital}
T.~Wysocki, H.~Razavi, and B.~Honary, \emph{Digital signal processing for
  communication systems}.\hskip 1em plus 0.5em minus 0.4em\relax Springer
  Science \& Business Media, 2013.

\bibitem{Zarrinkoub2014}
H.~Zarrinkoub, \emph{Understanding LTE with MATLAB: From Mathematical Modeling
  to Simulation and Prototyping}, 1st~ed.\hskip 1em plus 0.5em minus
  0.4em\relax Wiley Publishing, 2014.

\end{thebibliography}
\end{document}